\documentclass{article}

\usepackage{arxiv}
\usepackage[utf8]{inputenc} 
\usepackage[T1]{fontenc}    
\usepackage{hyperref}       
\usepackage{url}            
\usepackage{booktabs}       
\usepackage{amsmath}
\usepackage{amssymb}
\usepackage{amsfonts}       
\usepackage{nicefrac}       
\usepackage{microtype}      
\usepackage{graphicx}
\usepackage{caption}
\usepackage{subcaption}

\author{
    Chao Xu \\
    Department of Physics \\
    University of California, San Diego \\
    \AND    
    Jinyang Li, Tarek Abdelzaher, Heng Ji\\
    Department of Computer Science \\
    University of Illinois at Urbana Champaign \\
    \AND
    Boleslaw K. Szymanski \\
    Department of Computer Science \\
    Rensselaer Polytechnic Institute \\    
    \AND
    John Dellaverson\\
    Department of Computer Science \\
    University of California, Los Angeles \\
}

\newcommand\norm[1]{\left\lVert#1\right\rVert}

\title{The Paradox of Information Access: On Modeling Social-Media-Induced Polarization}

\begin{document}
\maketitle

\begin{abstract}
The paper develops a stochastic model of drift in human beliefs that shows that today's sheer volume of accessible information, combined with consumers' confirmation bias and natural preference to more outlying content, necessarily lead to increased polarization. The model explains the paradox of growing ideological fragmentation in the age of increased sharing.
As social media, search engines, and other real-time information sharing outlets purport to facilitate access to information, a need for content filtering arises due to the ensuing information overload. In general, consumers select information that matches their individual views and values. The bias inherent in such selection is echoed by today’s information curation services that maximize user engagement by filtering new content in accordance with observed consumer preferences. Consequently, individuals get exposed to increasingly narrower bands of the ideology spectrum, thus fragmenting society into increasingly ideologically isolated enclaves. We call this dynamic {\em the paradox of information access\/}. The model also suggests the disproportionate damage attainable with a small infusion of well-positioned misinformation. The paper describes the modeling methodology, and evaluates modeling results for different population sizes and parameter settings.  
\end{abstract}

\section{Introduction}
In human psychology, several well-known paradoxes exist when monotonicity of reward with respect to a beneficial stimulus is ultimately broken. For example, \emph{the paradox of choice} maintains that proliferation of choices ultimately leads to decreased satisfaction, as individuals perceive a higher opportunity cost in committing to their decisions~\cite{schwartz2004paradox}.

In this paper, we ask whether a different type of paradox exists in today's information age: Namely, does increased access to information (e.g., made possible by social media platforms that facilitate real-time publishing) lead to increased societal polarization? The idea that increased access creates polarization is not new. For example, it was shown that the creation of the interstate highway system in the US increased socio-economic disparity and geographic polarization in metropolitan areas~\cite{nall2015political}, as the ease of commute facilitated urban sprawl giving rise to homogeneous geographic neighborhoods (in an analogy with social echo-chambers) of significantly different character from downtown counterparts. 
We argue that similar observations apply to information access; the \emph{mere availability of more information ultimately increases fragmentation of society into ideologically isolated enclaves}. 

Informally, information access allows individuals to find and settle in ``ideological neighborhoods'' (echo-chambers) with other like-minded sources. Volume makes them less likely to explore other ideological neighborhoods. 
More specifically to the information landscape, public information available on social media platforms is intended (in theory) to have \emph{global visibility}. Yet, the resulting increased volume of information (created by a vastly enlarged number of diverse producers) makes \emph{human attention} an increasingly over-subscribed bottleneck. It necessitates information filtering, as recipients must make consumption choices.  This customized filtering, tailored to consumer biases (and aided by algorithmic curation services), essentially reinforces the bias, gradually eroding the common ground for dialogue among communities of different beliefs and creating the paradox: \emph{ideological fragmentation as a consequence of global access}.

The dynamics described above have significant societal consequences. In fact, one might go as far as saying that democracy itself, as conceived by modern society, is in danger. 
Democracy in modern society is based on the assumption that constituents are \emph{well-informed}. In an environment, where selective exposure and information disorders distort reality, the foundations of democracy themselves may be jeopardized. 

\begin{figure*}[thb]
    \centering
    \includegraphics[width=0.8\linewidth]{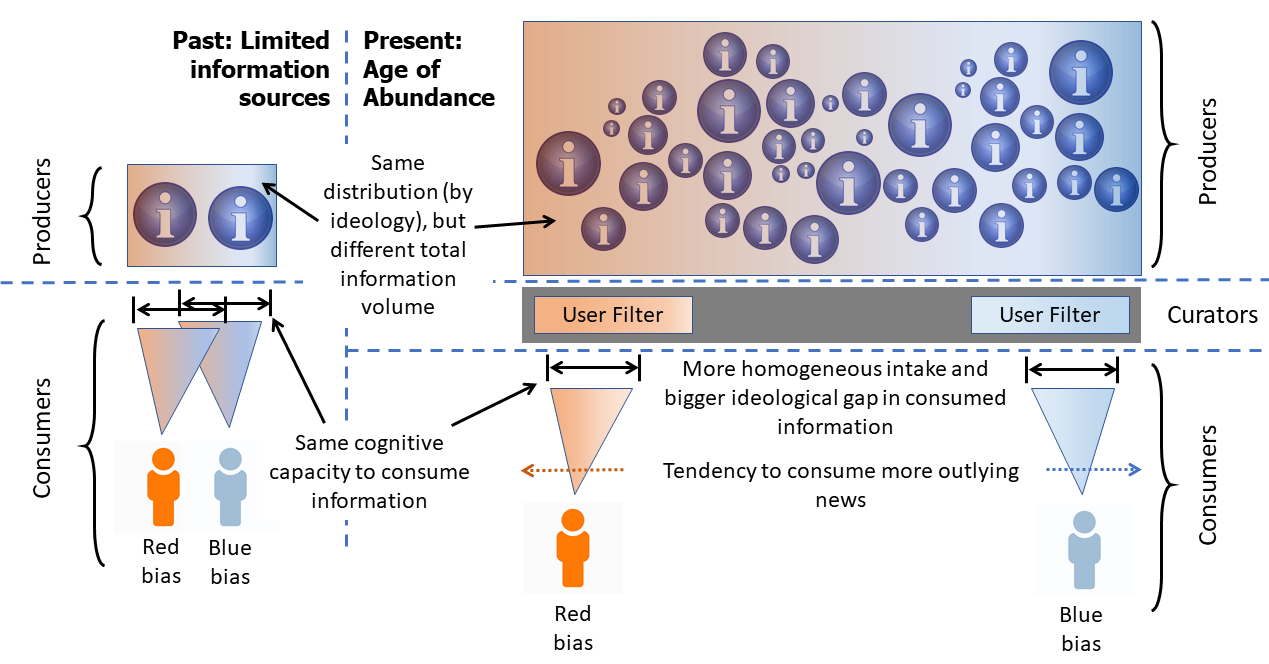}
    \caption{Notional illustration of the impact of information volume on the evolution of bias. Assuming that information curation services will recommend information that accurately matches consumer bias, given the same amount of consumed information, content given to a particular consumer will be much more homogeneous (on the ideology spectrum), and content given to different consumers may have bigger ideological gaps, compared to a situation with more limited sources.}
    \label{fig:gap}
\end{figure*}

In the rest of this paper, we develop a model that shows how the mere increase in the volume of accessible information is sufficient to create polarization, assuming the existence of confirmation bias~\cite{nickerson1998confirmation} and consumer preference for more outlying content~\cite{fiske1980attention,shoemaker1996hardwired,lamberson2018model,varshney2019must} (both of which are traits borrowed from social psychology). Figure~\ref{fig:gap} notionally demonstrates this effect. 
When this volume increases, our individual \emph{coverage} of available information decreases (i.e., the percent of it that an individual can consume). Larger ideological gaps emerge between information consumed by different parties, leading to polarization.
  
\section{The Paradox of Information Access}
Our model represents consumers as particles in a spherical belief space, where distance from origin denotes departure from neutrality (whereas the angle denotes the particular bias). Unlike the early days of broadcast, where reaching a large audience needed specialized resources, in the age of social media, anyone can be a content producer. We call this model {\em democratized production\/}. Democratized production implies that costumers are also  producers. Therefore, the distribution of generated 
information in the belief space is proportional to the population density in that space. The evolution of a node's belief is then modeled by a motion equation. This motion (in the belief space) is driven by the density of surrounding nodes plus some noise, giving rise to a particle diffusion-drift model. We show in this section that the model reveals a   bifurcation of population density in the belief  space (i.e., emergence of polarization). The bifurcation is more pronounced when overload increases. 

More specifically, let $x_{i, t} \in \mathbb{R}^K$ denote the position of individual $a_i~(i=1, 2,\cdots, n)$ in a K-dimensional ``belief space'' at time \(t\). It reflects their beliefs at time \(t\). We assume the origin point of the belief space is neutrality. Departures from the origin represent bias towards specific beliefs, where {\em direction\/} encodes the nature of the belief (e.g., liberal versus conservative) and {\em magnitude\/} represents the degree to which the belief is upheld, where $0$ is neutral. Let
$x_{i, t}$ reflect how extreme agent \(a_i\)'s belief is at time $t$. Let \(\mathbf{X_t}\) denote the entire population of \(n\) agents. Similarly, we define the set of documents, $\mathbf{Y}_t$, where $y_{j,t} \in \mathbf{Y}_t$ represents the belief espoused by the $j$th document, produced at time $t$.




\subsection{Democratized Production}

We assume that, on a fully democratized medium, everyone is a content producer who propagates its current belief. Thus:
\begin{equation}
    \mathbf{Y}_t = \mathbf{X}_t
    \label{eq:production}
\end{equation}
In other words, we shall henceforth use $x_{i,t}$ and $y_{i,t}$ interchangeably to denote both the position of an agent and the position of the content they generate. Clearly, this is an over-approximation, but we simplify in the interest of tractability. 

\subsection{Curation and Confirmation Bias}

Aside from other random influences, we assume that an individual, $a_i$, of position $x_{i,t}$ at time $t$, will engage with a subset of content, $\mathbf{Y}^{(i)}_t$, that matches their belief; a phenomenon known as {\em confirmation bias\/}~\cite{nickerson1998confirmation}. This subset will be preferentially sought by the individual and thus also preferentially ranked by the curator. For each consumer, we assume that a visibility radius $\epsilon$ limits how ideologically distant the content they engage with might be. Thus:
\begin{equation}
    \mathbf{Y}^{(i)}_t = \{y_{j,t} | y_{j,t} \in \mathbf{Y}_t,  \|x_{i,t} - y_{j,t}\| \leq \epsilon\}
    \label{eq:curator}
\end{equation}

In an age of overload, an inherent assumption is that $\epsilon$ is small compared to (the radius of) the overall belief space. This allows linearization-based approximations within radius, $\epsilon$, that facilitate analysis. We call $\mathbf{Y}^{(i)}_t$ consumer $i$'s \emph{neighborhood set}.


\subsection{Consumer Preference for Outlying Content}
\label{sec:outlying}
Modern information propagation behaviors are a combination of ingrained biases and efficiency shortcuts, exacerbated by overload. For example, we increasingly seek (and spread) more sensational and surprising news. Seeking (and spreading) more outlying or surprising news~\cite{lamberson2018model,varshney2019must} is a collective {\em learning efficiency\/} tactic in that it avoids expending cognitive resources on processing redundant inputs. 
The tactic seems increasingly pronounced as the volume of information increases~\cite{varshney2019must}, arguably biasing our collective attention towards more extreme content. Thus, 
consistently with prior literature~\cite{fiske1980attention,shoemaker1996hardwired,lamberson2018model,varshney2019must}, we take into account that, of the content consumed by an agent, more outlying news have a deeper influence. Accordingly, a consumed item in $\mathbf{Y}^{(i)}_t$, that lies at location $x$ in the belief space, has an influence weight $\eta_0(x)$ that generally increases with distance $\norm{x}$ from the (neutral) origin. However, beyond a certain $\norm{x}$, influence decreases again, when the espoused beliefs become ``too extreme''.

\subsection{Social Influence}
In the age of information overload, following social influence is an efficiency tactic~\cite{bonabeau2004perils}. Replicating behaviors of role models and like-minded individuals obviates expending one's own cognitive resources (e.g., on ascertaining veracity of information we forward). Evidence suggests that its usage becomes more pronounced under time pressure~\cite{buckert2017imitation}. 
Social influence works super-linearly. That is to say, it rises sluggishly at first, then as the number of neighboring adopters increases, it rises progressively faster. We model this notion by assuming that a consumed item in $\mathbf{Y}^{(i)}_t$, that lies at location $x$ in the belief space, has a component of influence that increases exponentially with the population density around $x$. Let the \emph{density} of content items at location $x$ and time, $t$, be denoted by $\rho (x, t)$. Thus, content influence increases with $e^{\kappa \rho (x,t)}$. The constant $\kappa$ describes the relative impact of social influence. If $\kappa=0$, this factor is neutralized (i.e., becomes identically 1). In this case, each individual item contributes an independent influence regardless of its agreement with other items; the influence of item collections grows {\em linearly\/} with collection size. Otherwise, if $\kappa > 0$, the higher the $\kappa$, the more rapid the (super-linear) escalation of influence of items (around location $x$) with density of adoption of $x$. (As we show later, polarization emerges even with $\kappa = 0$, but it increases with $\kappa$.)

Taking both (i) preference for outlying content (from Section~\ref{sec:outlying}) and (ii) social influence into account, the influence weight of an item at location $x$ within consumer $i$'s neighborhood set, $\mathbf{Y}^{(i)}_t$, is given by:

\begin{equation}
\eta_t (x) = \eta_0 (x) e^{\kappa \rho (x,t)}
\label{eq:weights}
\end{equation}
where $\eta_0(x)$ increases with distance $\norm{x}$ from the neutral origin (to indicate preference for outlying content) up to a point, until the beliefs become too extreme, then declines. 

\subsection{A Belief Update Model}

We are now ready to examine how content consumption affects individual beliefs. Let us examine the $i^{th}$ consumer. Upon consuming information, the consumer is attracted towards the (ideological) center of gravity, $\bar{y}_{i,t}$, of consumed information items, each weighted by their influence upon the consumer. Let $\Delta W$ be the cumulative effect of all other random influences on the consumer's position within time $\Delta t$. These may include effect of random social contacts (long links) and effect of content that lies outside the consumer's neighborhood set. Assuming that individual displacements produced by these random factors constitute white noise, the cumulative effect, $\Delta W$, is the integral of white noise over $\Delta t$, which is given by a Weiner process (i.e., Brownian motion).
Thus, the equation of motion (or the belief update)
of this consumer is given by:
\begin{align}
    x_{i,t+\Delta t} &= (1 - \alpha_{\Delta t}) x_{i,t} + \alpha_{\Delta t}\bar{y}_{i,t} +
    \sigma \Delta W\\
    x_{i,t+\Delta t} &= x_{i,t} + \alpha_{\Delta t}(\bar{y}_{i,t} - x_{i,t})
    +\sigma \Delta W\,.
    \label{eq:update}
\end{align}
In the above equation, 
$\alpha_{\Delta t}$ is a constant, $0 \leq \alpha_{\Delta t} \leq 1$, that represents how ``impressionable'' an individual is. A larger $\alpha_{\Delta t}$ leads to a larger belief update during time ${\Delta t}$. In addition,
$\sigma$ is constant that scales the influence of extraneous factors on the belief update. A particularly ``narrow'' individual, for example, might be primarily influenced by their immediate neighbors, and thus have a low $\sigma$. 

\begin{figure}[htb]
    \centering
    \includegraphics[width=0.95\linewidth]{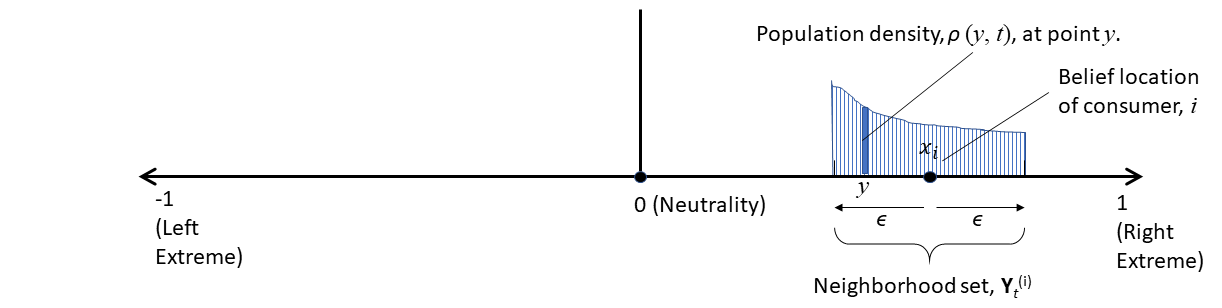}
    \caption{An illustration of one dimensional beliefs.}
    \label{fig:belief}
\end{figure}

Let us first consider motion in one dimension only; namely, the radial dimension (i.e., motion that is either away from neutrality or towards neutrality). This one-dimensional case is shown in Figure~\ref{fig:belief}. We can estimate $\bar{y}_{i,t}$ by integrating over the neighborhood $\epsilon$ (around the consumer's position, $x$) the positions of all consumed items. Remember that the item density at position $y$ is $\rho (y, t)$, and that each item has influence (weight), $\eta_t(y)$, given by Equation~(\ref{eq:weights}). Thus:

\begin{equation}
    \bar{y} = \frac{\int_{x-\epsilon}^{x+\epsilon} \rho(y,t) \eta_t(y) y\,dy}{\int_{x-\epsilon}^{x+\epsilon} \rho(y,t) \eta_t(y)\,dy}
    \label{eq:y}
\end{equation} 

Since $\epsilon$ is small, we can approximate the functions $\rho(y,t)$ and $\eta_t(y)$ by their first order Taylor series expansion. Thus: 
    
\begin{align}
    \bar{y} &\simeq \frac{\int_{x-\epsilon}^{x+\epsilon} (\rho(x,t) + \rho^{\prime}(x,t)(y-x))(\eta_t(x) + \eta^{\prime}_t(x)(y-x))y dy}{\int_{x-\epsilon}^{x+\epsilon} (\rho(x,t) + \rho^{\prime}(x,t)(y-x))(\eta_t(x) + \eta^{\prime}_t(x)(y-x))dy}\\
    &\simeq \frac{\rho(x,t)\eta_t(x)\cdot 2x\epsilon + (\rho(x,t)\eta^{\prime}_t(x)+\rho^{\prime}(x,t)\eta_t(x))(2\epsilon^3/3)}{\rho(x,t)\eta_t(x)\cdot 2\epsilon }\\
    &= x + \frac{\epsilon^2}{3}(\eta^{\prime}_t(x)/\eta_t(x)+\rho^{\prime}(x,t)/\rho(x,t))\,.
\end{align}

Substituting from the above equation into Eq.~\eqref{eq:update}, the belief update can then 
be formulated as:
\begin{equation}
    x_{i,t+\Delta t} - x_{i,t} \simeq (\alpha_{\Delta t}) \frac{\epsilon^2}{3}(\eta^{\prime}_t(x_{i,t})/\eta_t(x_{i,t})+\rho^{\prime}(x_{i,t},t)/\rho(x_{i,t},t)) + \sigma \Delta W\,.
    \label{eq:motion}
\end{equation}

Without losing generality, we consider the case where $x_{i,t} > 0$. The above equation suggests that (barring the noise term) individuals will generally move towards more extreme beliefs (i.e., $x_{i,t+\Delta t} > x_{i,t}$), unless $\rho^{\prime}(x_{i,t},t)/\rho(x_{i,t},t) < - \eta^{\prime}_t(x_{i,t})/\eta_t(x_{i,t})$. In other words, the belief density, $\rho(x_{i,t},t)$, should ideally have a sufficiently negative slope (fewer adopters of more outlying beliefs) to the right of the neutrality point to meet the above condition. Otherwise, polarization occurs. Below, we study these dynamics more formally.

Dividing  Equation~(\ref{eq:motion}) by $\Delta t$ and taking the limit of as $\Delta t \to 0$, we get:
\begin{equation}
    \frac{d x_{i,t}}{dt} \simeq \mu (\ln \eta_t(x_{i,t}))^\prime  + \mu \rho^{\prime}(x_{i,t},t)/\rho(x_{i,t},t)
    +\sigma\frac{dW}{dt}\,,
    \label{eq:eom}
\end{equation}
where
\begin{equation}
    \mu = \frac{\epsilon^2}{3}\lim_{\Delta t \to 0} \frac{\alpha_{\Delta t}}{\Delta t}  \,.
    \label{eq:mu}
\end{equation}

Referring back to Equation~(\ref{eq:update}), it is interesting to observe that $\lim_{\Delta t \to 0} \frac{\alpha_{\Delta t}}{\Delta t}$ is the normalized rate of change of consumer's belief (specifically, the rate of change of consumer's belief per unit of distance from the center of gravity of content they consume). The above equation represents dynamics of population's beliefs. In the continuity limit, it happens to have an equivalent diffusion-drift description. Eq.~\eqref{eq:eom} is the generalized Fokker-Planck equation~\cite{fokker1914mittlere,planck1917satz}, where $\frac{d x_{i,t}}{dt}$ is the instant velocity of each individual particle; $\mu$ is the drift coefficient, which describes the mobility of each one; $\sigma$ is the intensity a standard Wiener process; $\rho(x_{i,t},t)$ is the probability density in the context of stochastic process.

\subsubsection{Diffusion-Drift Description}
Let us substitute into Equation~(\ref{eq:eom}) for $\eta_t(x)$ from Equation~(\ref{eq:weights}). 
Recognizing that Eq.~\eqref{eq:eom} is a hydro-dynamical flow equation, we can thus write down the local population flow
$j(x)$ as (see Appendix A for derivation detail)~\cite{klyatskin2010lectures}: 
\begin{equation}
    j(x,t) = -D\partial_x \rho(x,t) - \mu \rho(x,t) \partial_x V(x) - \mu \rho(x,t)\partial_x (g\rho(x,t))\,.
    \label{eq:j}
\end{equation}

In the above equation, according to the hydro-dynamic analogy, $\rho(x,t)$ is the local density of the compressible fluid;
$D = \sigma^2/2 - \mu$ is the diffusion constant, 
$V(x) = -\ln{\eta_0(x)}$ is an external potential applied to the fluid (fluids tend to
converge to positions with relatively lower potential), 
and $g = -\kappa$, which can be interpreted as a strength of ``attraction" among the particles. 
At equilibrium, we should approach a steady state, where:
\begin{equation}
    \frac {\partial j(x,t)}{\partial x} = 0\,.
    \label{eq:steady}
\end{equation}

Taking the partial derivative of Equation~(\ref{eq:j}) with respect to $x$, and applying Equation~(\ref{eq:steady}), we get an equation that describes the equilibrium distribution of population density, $\rho$:
\begin{equation}
    D\partial_x^2 \rho + \mu\partial_x( \rho\partial_x V)  + \frac{1}{2}g\mu
    \partial_x^2(\rho^2)= 0\,,
\end{equation}
which is a generalized diffusion-drift equation~\cite{landau1987theoretical} containing a non-linear effect. Solving the above equation for $\rho$ computes the steady state outcome (namely, density distribution) according to the modeled social dynamics. Of specific interest in this case is to determine if the steady state distribution is bimodal (i.e., polarized) or not. This is explored in the figures shown next. 

\subsection{Model Simulation Results}
\begin{figure}
    \centering
    \begin{subfigure}[b]{0.49\textwidth}
        \centering
        \includegraphics[width=\textwidth]{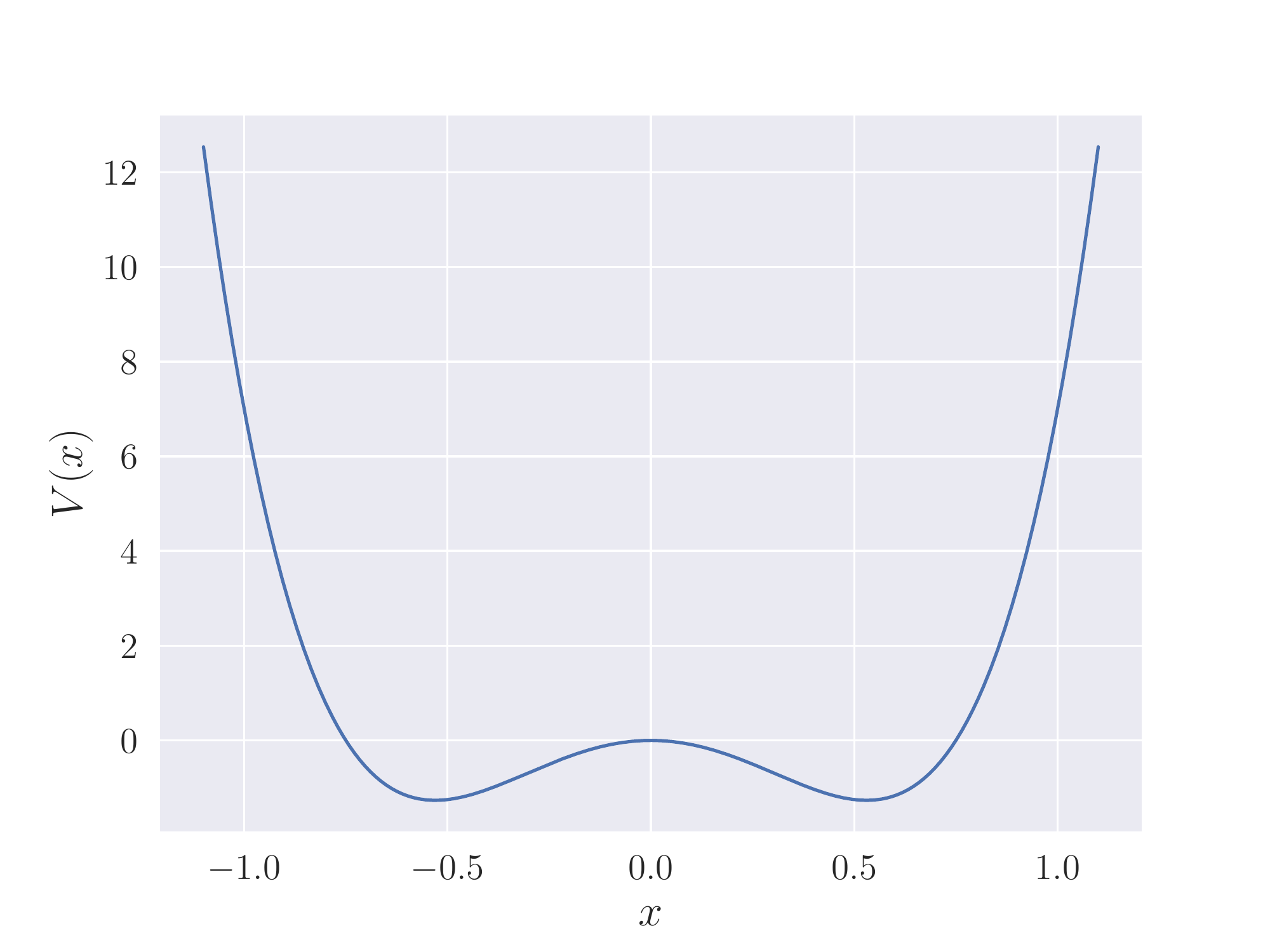}
        \caption{V(x)} 
        \label{fig:vx}
    \end{subfigure}
    \hfill
    \begin{subfigure}[b]{0.49\textwidth}
        \centering
        \includegraphics[width=\textwidth]{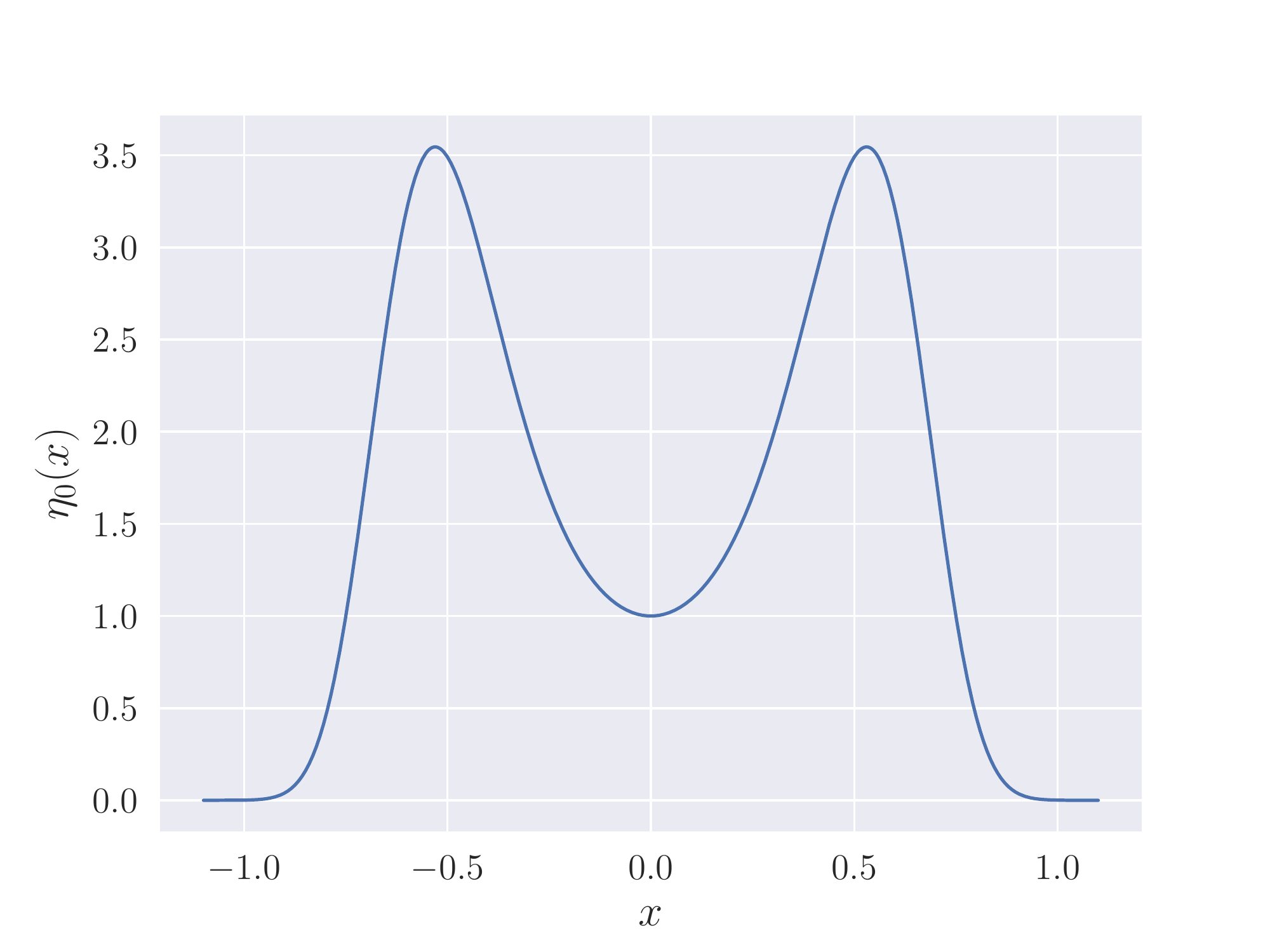}
        \caption{$\eta_0(x) = e^{-V(x)}$}
        \label{fig:eta0}
    \end{subfigure}
    \caption{Potential used in the simulation}
    \label{fig:potential}
\end{figure}

The above model can be simulated to describe the evolution of belief in societies with ubiquitous production and consumer-driven curation. In the simulation, we consider a one-dimensional case where population in the belief space can bifurcate, resulting in 2 peaks, or not, resulting in one peak. The used $\eta_0(x)$ and corresponding $V(x)$ are visualized in Figure~\ref{fig:potential}. This potential $V(x)$ is soft, meaning that people exhibit a mild tendency to be more affected by more extreme content, pushing them away from the neutral position $x=0$. Yet, even such a soft potential is still able to cause bifurcation if people are overloaded.


The bifurcation is measured as visibility of the two potential bifurcated peaks:
\begin{equation}
    Q = \frac{peak - valley}{peak + valley}
\end{equation}
If bifurcation occurs, then the visibility $Q$ approaches 1 as the degree of bifurcation become higher. If there is no bifurcation, then visibility $Q = 0$.

\begin{figure}[!htb]
    \centering
    \begin{subfigure}[b]{0.49\textwidth}
        \centering
        \includegraphics[width=\textwidth]{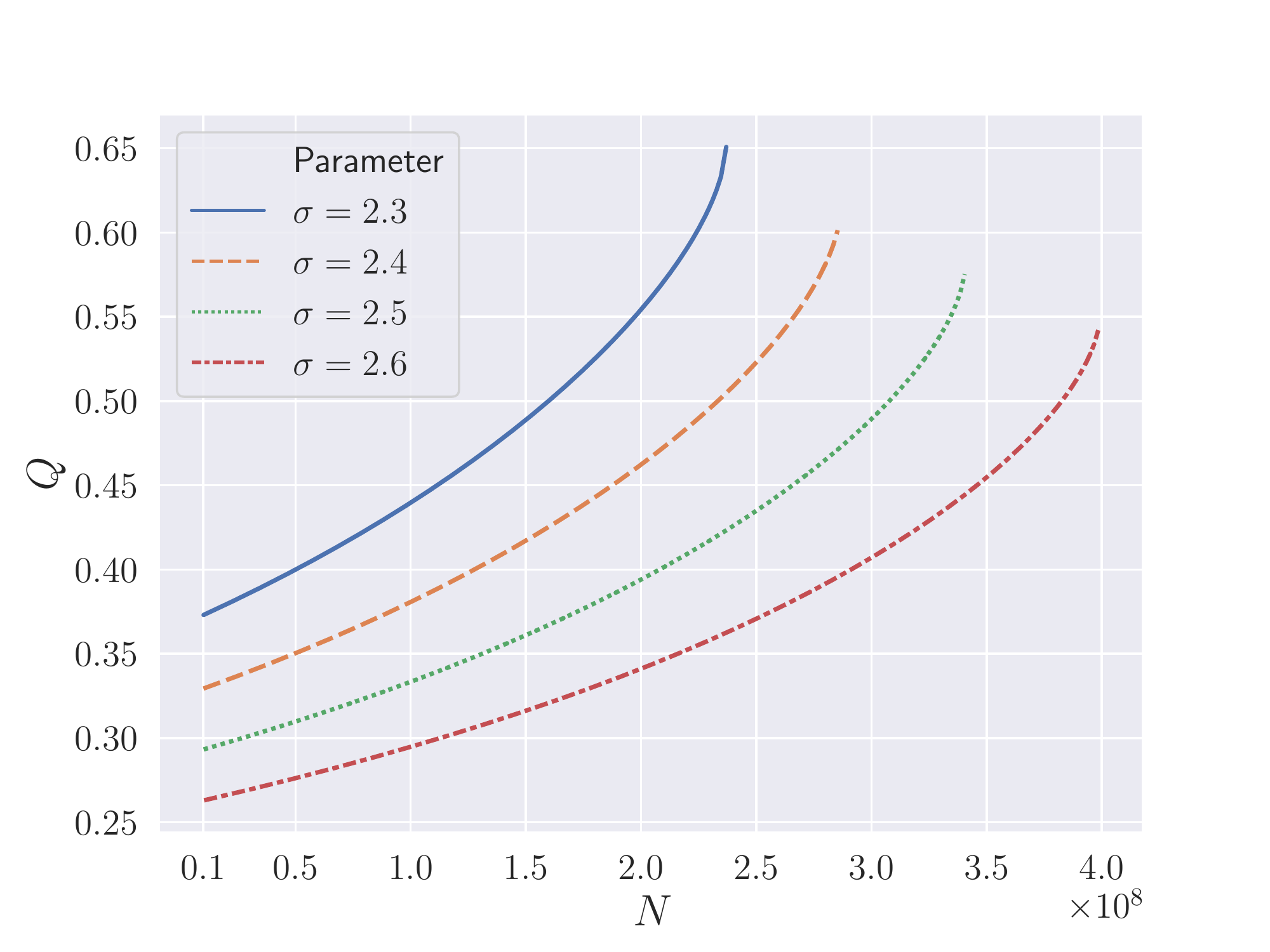}
        \caption{$\mu=1.0$, $\kappa=5.0\mathrm{E}{-9}$}
        \label{fig:q-n-sigma}
    \end{subfigure}
    \hfill
    \begin{subfigure}[b]{0.49\textwidth}
        \centering
        \includegraphics[width=\textwidth]{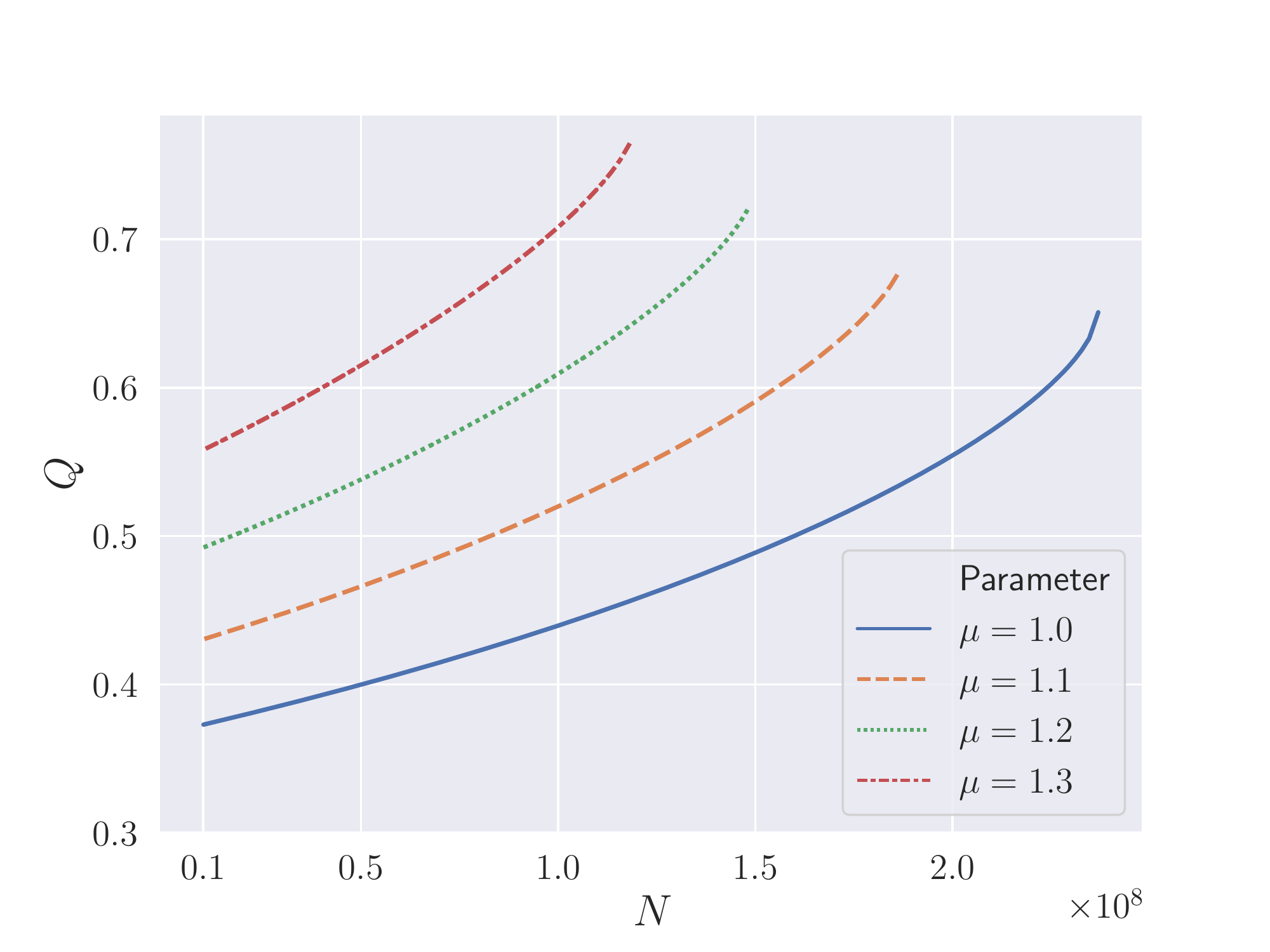}
        \caption{$\sigma=2.3$, $\kappa=5.0\mathrm{E}{-9}$}
        \label{fig:q-n-mu}
    \end{subfigure}
    \caption{Population bifurcation as a function of total population $N$. Here $N$ reflects how overload the system is. As $N$ increases, overload increases and bifurcation becomes more visible.}
    \label{fig:q-n}
\end{figure}

\begin{figure*}[!htb]
    \centering
    \includegraphics[width=0.5\linewidth]{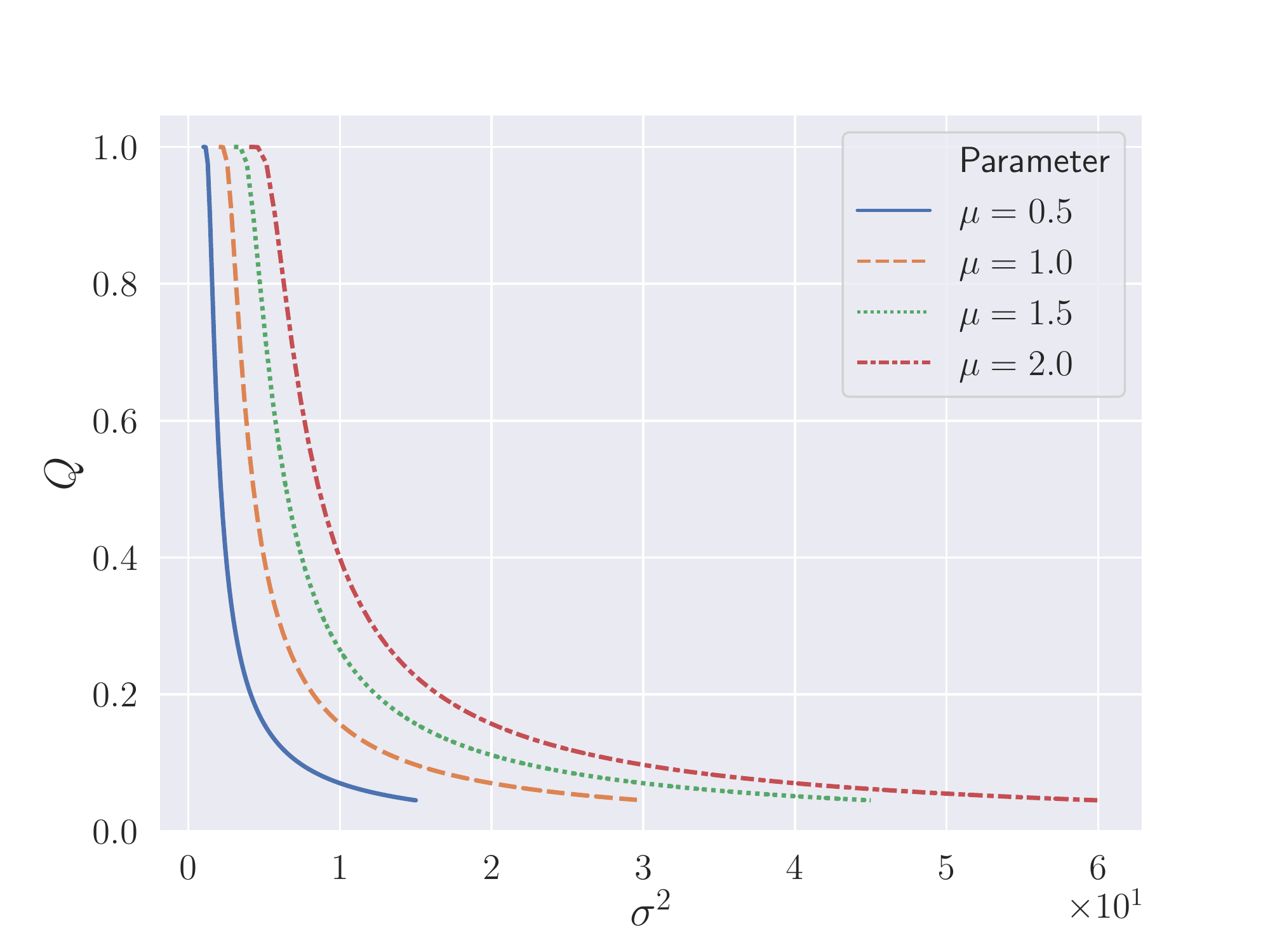}
    \caption{Population bifurcation as a function of diversity (randomness) in the content consumed. As diversity $\sigma$ increases, bifurcation becomes less visible. $\kappa = 0.0$, $N = 1.0\mathrm{E}{8}$.}
    \label{fig:q-sigma}
\end{figure*}

\begin{figure}[!htb]
    \centering
    \begin{subfigure}[b]{0.51\textwidth}
        \centering
        \includegraphics[width=\textwidth]{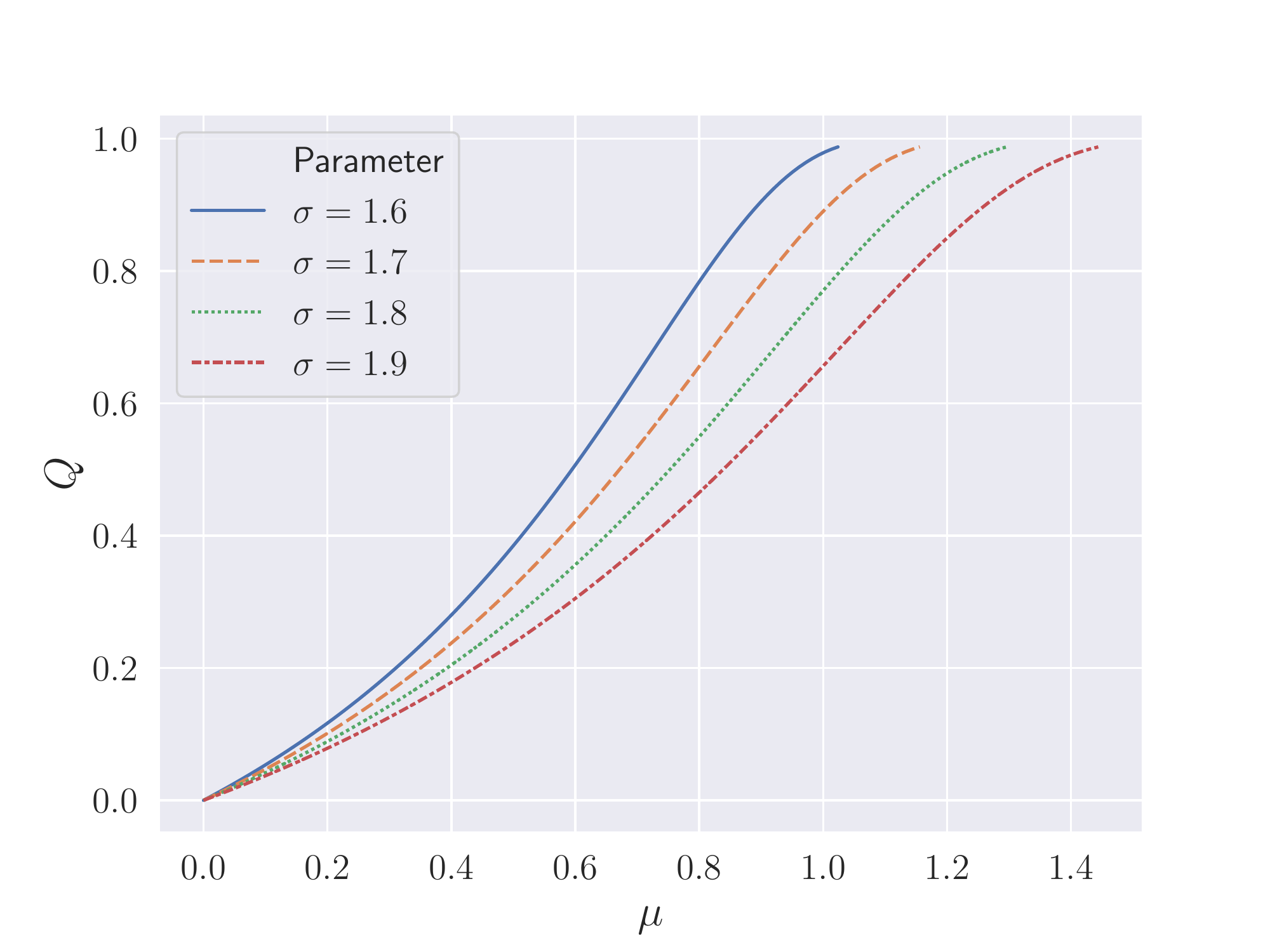}
        \caption{$\kappa = 0.0$, $N = 1.0\mathrm{E}{8}$}
    \end{subfigure}
    \hfill
    \begin{subfigure}[b]{0.49\textwidth}
        \centering
        \includegraphics[width=\textwidth]{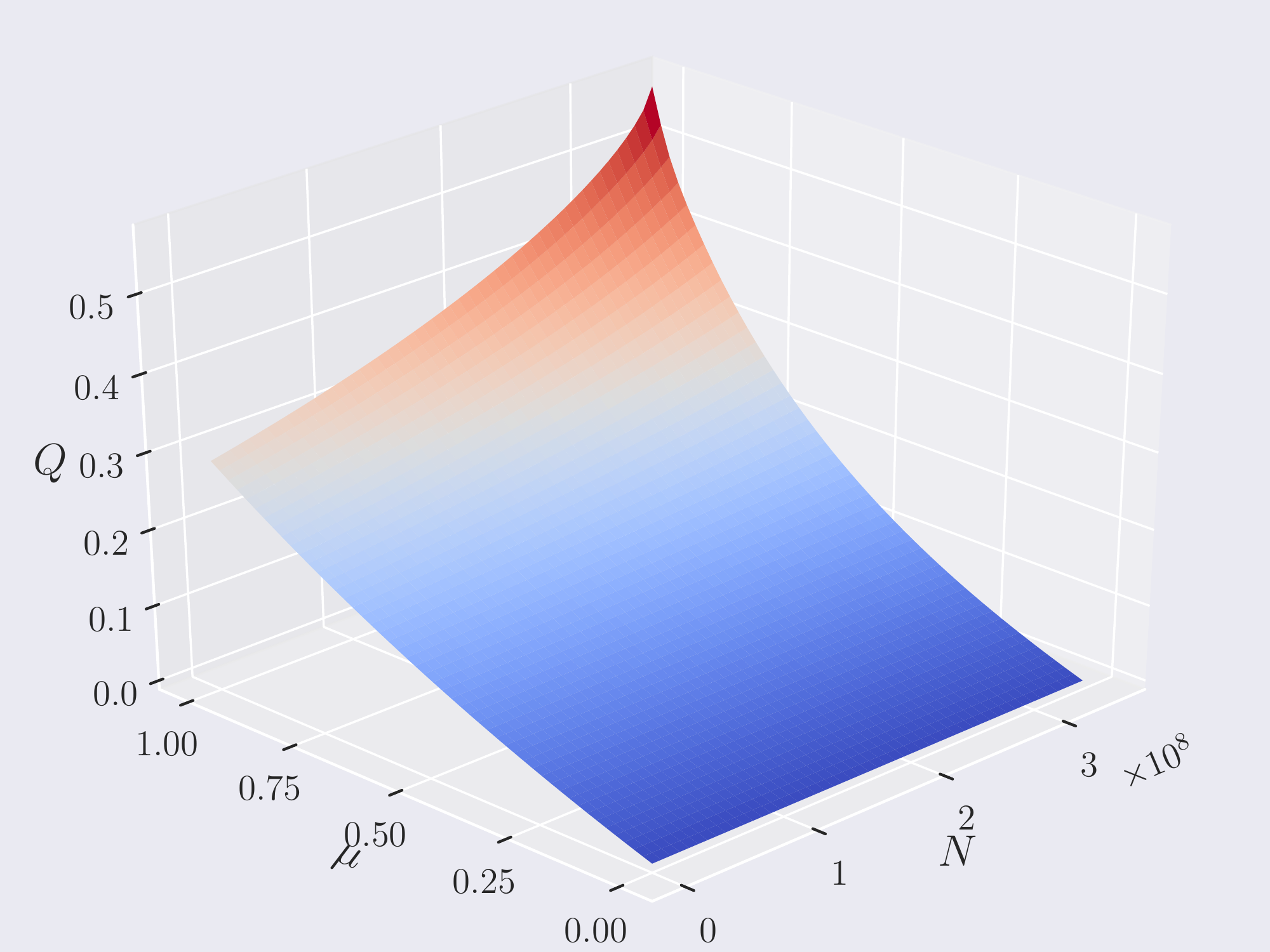}
        \caption{$\sigma=2.5$, $\kappa = 5.0\mathrm{E}{-9}$.}
    \end{subfigure}
    \hfill
    \begin{subfigure}[b]{0.49\textwidth}
        \centering
        \includegraphics[width=\textwidth]{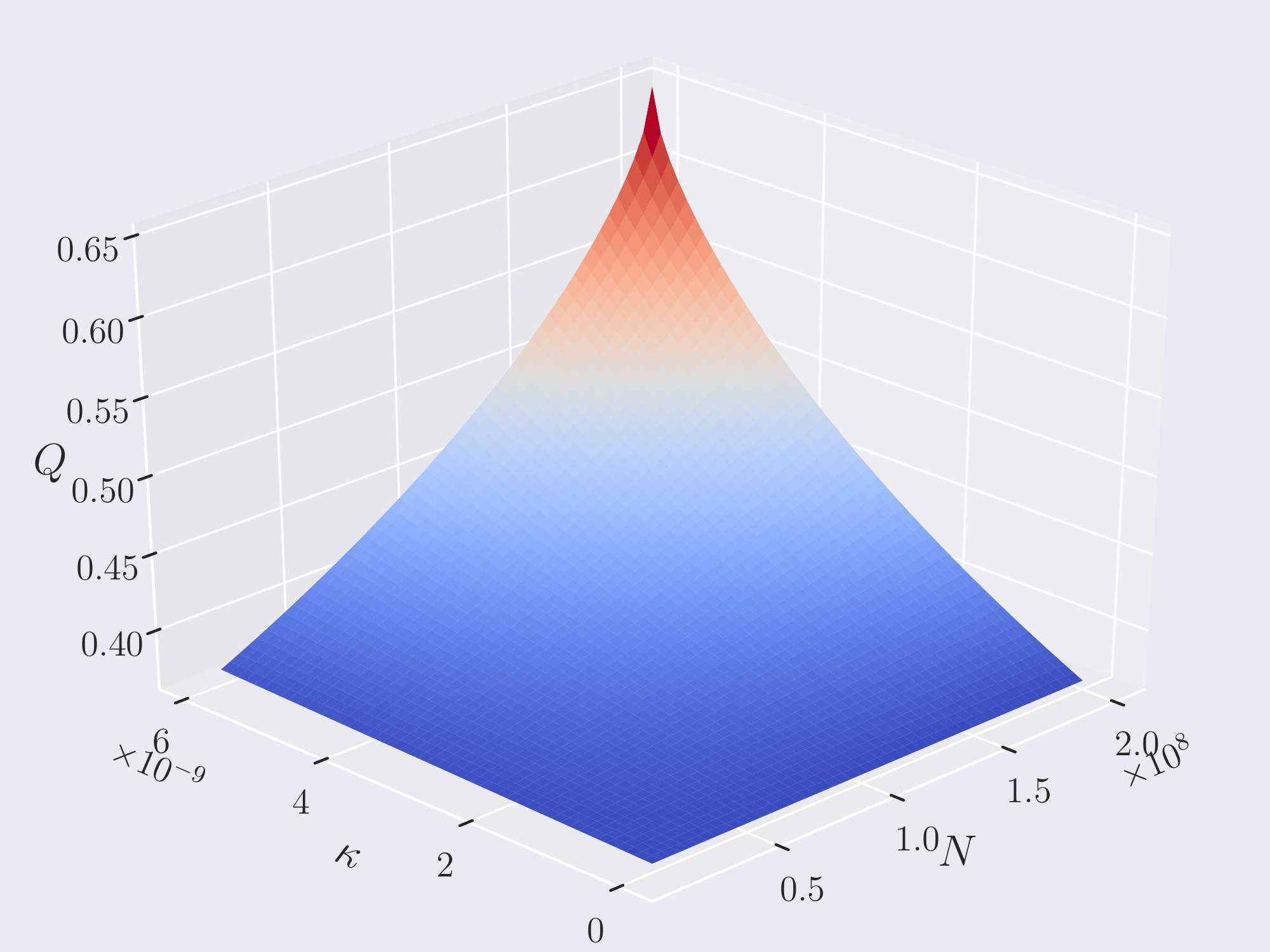}
        \caption{$\sigma=2.3$, $\mu = 1.0$.}
    \end{subfigure}
    \caption{Population bifurcation as a function of impressionability $\mu$. As impressionability $\mu$ increases, bifurcation becomes more visible.}
    \label{fig:q-mu}
\end{figure}

A few trends can be observed:

\begin{itemize}
    \item \emph{Effect of overload:} Figure~\ref{fig:q-n} shows the bifurcation of population density in the belief space changing as the amount of content, $N$, increases. As shown in the figure, as the volume of information increases, bifurcation becomes higher ($Q \to 1$). The effect is exacerbated when the impressionability, $\mu$, increases, or when the impact of other orthogonal and diverse influences, $\sigma$, decreases.

    \item {\em Effect of diversity:\/} Figure~\ref{fig:q-sigma} shows how the diversity of influence affects the population's belief distribution. Parameter $\sigma$ describes the degree to which belief updates are affected by random (i.e., diverse) factors outside the immediate belief neighborhood of consumers. As expected, increasing the $\sigma$ (and thus decreasing the relative impact of confirmation bias) has a beneficial effect.
   
    \item \emph{Effect of impressionability:} More impressionable consumer populations have a higher rate of belief change, $\alpha$, and thus, $\mu$, for the same distribution of neighboring content items. It can be seen from Figure~\ref{fig:q-mu} that as the individual's impressionability increases, the degree of bifurcation becomes higher ($Q \to 1)$.
\end{itemize}

The figures confirm that the mere increase in content volume leads to increased polarization. The effect is further magnified by social influence, $\kappa$, and impressionability, $\mu$, although it exists even if $\kappa = 0$. Below, we show that content injected in appropriately chosen parts of the belief space can further have disproportionate impact on increasing polarization.

\section{Committed Radical Communities, and Misinformation} 
Consider the case where some \emph{small} subset of producers are in fact bots or hard-line groups committed to outlying beliefs. By committed, we mean that the positions of hard-liners in the belief space do not change over time, regardless of surrounding influence. Let $\rho_H (x)$ represent their population density in the belief space. Note that, since all positions are fixed, the density is also fixed and is not a function of time.

In this enhanced model, belief updates for the remaining population, originally given by Equation~(\ref{eq:update}), are now given by:

\begin{equation}
    x_{i,t+\Delta t} = x_{i,t} + \alpha_{\Delta t}(\bar{y}^E_{i,t} - x_{i,t})
    +\sigma \Delta W\,
    \label{eq:update2}
\end{equation}

where $\bar{y}^E_{i,t}$, the new version of $\bar{y}_{i,t}$ (denoting the center of gravity towards which each particle drifts), is rewritten to account for the effect of $\rho_H (x)$ as well. Equation~(\ref{eq:y}) is thus replaced with the following:

\begin{equation}
    \bar{y}^E = \frac{\int_{x-\epsilon}^{x+\epsilon} (\rho(y,t) + \rho_H(y))\eta_t(y) y\,dy}
    {\int_{x-\epsilon}^{x+\epsilon} (\rho(y,t) + \rho_H(y)) \eta_t(y)\,dy} 
    \label{eq:yy}
\end{equation} 

Similarly, Equation~(\ref{eq:weights}) becomes:

\begin{equation}
    \eta^E_t(x) = \eta_0(x)e^{\kappa (\rho(x,t) + \rho_H(x))}\,.
    \label{eq:weight2}
\end{equation}

Following the same steps as before, Equation~(\ref{eq:eom}) becomes:

\begin{equation}
    \frac{d x_{i,t}}{dt} \simeq \mu (\ln \eta^E_t(x_{i,t}))^\prime  + \mu (\rho^{\prime}(x_{i,t},t)+\rho^\prime_H)/(\rho(x_{i,t},t) + \rho_H(x_i))
    +\sigma\frac{dW}{dt}\,.
    \label{eq:eom2}
\end{equation}

Assuming that $\rho_H(x) \ll \rho(x,t)$, we can re-write:

\begin{equation}
    \frac{d x_{i,t}}{dt} \simeq \mu (\ln \eta^E_t(x_{i,t}))^\prime  + \mu (\rho^\prime(x_{i,t},t)+\rho^\prime_H)/\rho(x_{i,t},t)
    +\sigma\frac{dW}{dt}\,.
    \label{eq:eom3}
\end{equation}

\begin{figure}[!htb]
    \centering
    \begin{subfigure}[b]{0.49\textwidth}
        \centering
        \includegraphics[width=\textwidth]{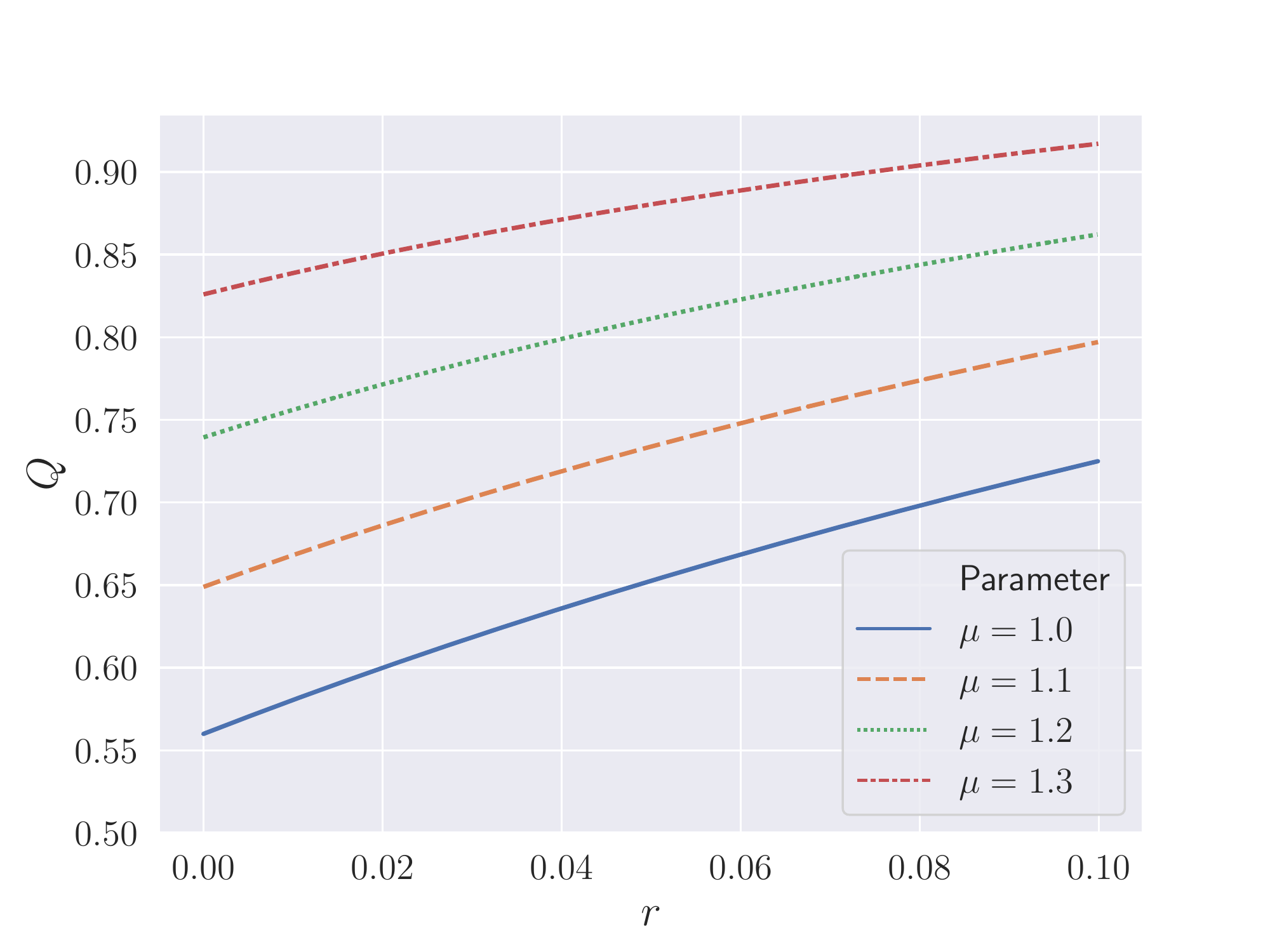}
        \caption{$\kappa = 0.0$, $\sigma=2.0$, $N = 1.0\mathrm{E}{8}$}
    \end{subfigure}
    \hfill
    \begin{subfigure}[b]{0.49\textwidth}
        \centering
        \includegraphics[width=\textwidth]{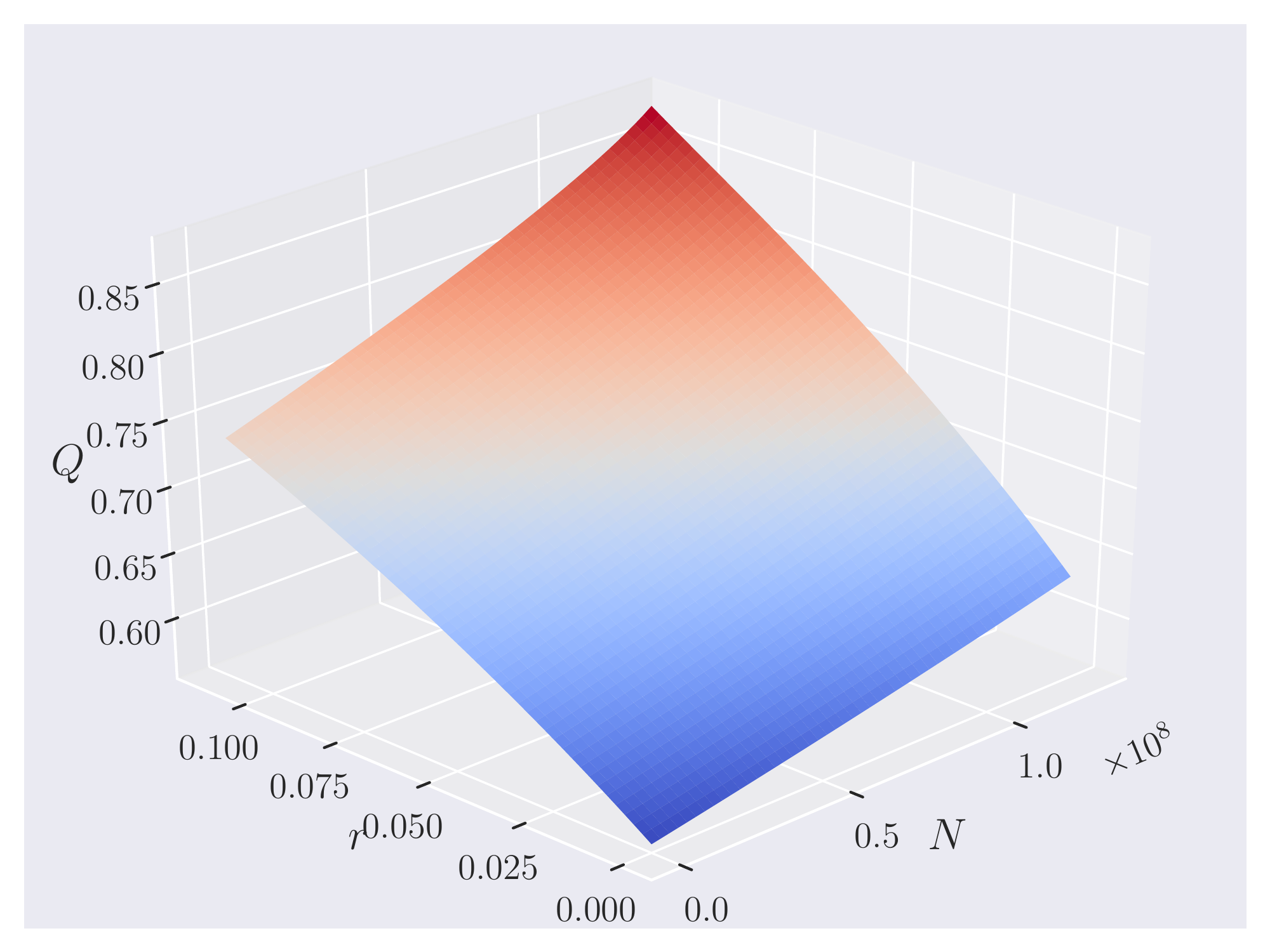}
        \caption{$\sigma=2.0$, $\mu=1$, $\kappa=2.0\mathrm{E}{-9}$.}
    \end{subfigure}
    \caption{Bifurcation of normal population as a function of the percentage of committed hard-liners over the total population. $r=N_\text{committed}/(N_\text{normal}+N_\text{committed})$}
    \label{fig:q-committed}
\end{figure}

Continuing with the derivation similarly to the previous section, we ultimately get the equilibrium equation:

\begin{equation}
    D\partial_x^2 \rho + \mu\partial_x( \rho\partial_x V^E)  + \frac{1}{2}g\mu
    \partial_x^2(\rho^2)= \mu \partial^2 \rho_H\,,
\end{equation}

where $V^E = - (\ln\eta_0 - g \rho_H)$, and (as before) $g = -\kappa$, $D = \sigma^2/2 - \mu$, and $\mu$ is given by Equation~(\ref{eq:mu}). 
We model the committed radical communities by adding a fixed distribution of $\rho_H(x)$, that
are two Gaussians located at the two maxima of $\eta_0(x)$, with standard deviation 
$\sigma_0^2 = 0.001$ in the belief space. 
When the ratio between the hard-liners and the total population increases, as expected,
bifurcation is significantly exacerbated as shown in Fig.~\ref{fig:q-committed}, especially when volume $N$ is high.

\section{Related Work}
The emergence of polarization as a result of local interaction and conversion among individuals with similar opinions has been studied extensively for decades~\cite{axelrod1997dissemination,li2013consensus,klemm2005globalization}. It is widely understood that the preference to interact with more similar individuals creates polarization~\cite{klemm2005globalization}. It is also observed that polarization leads to selective exposure to media (preferring media with a closer ideological match)~\cite{iyengar2009red}, and that polarization breeds misperceptions, even in the presence of corrective information~\cite{flynn2017nature}. 

Research in the early days of Internet communication suggests that computer-mediated communication itself may contribute to polarization~\cite{sia2002group}. Evidence was also reported that social media potentially exacerbate selective exposure~\cite{bakshy2015exposure} and that selective exposure and polarization may be mutually reinforcing~\cite{stroud2010polarization}. The role of algorithmic curation has been the subject of more debate. Some evidence suggests that algorithmic curation services contribute to selective exposure, resulting in filter bubbles~\cite{pariser2011filter}. Other models qualify this trend by suggesting it is true only when absence of curation does not materially reduce quality of content~\cite{berman2020curation}. Some research indicates that the effect of customized algorithmic curation on selective exposure is higher than the effect of user-driven bias~\cite{dylko2017dark}. Other work shows that it is unclear whether user-driven biases or curator-driven belief reinforcement (filter bubbles) have more impact on selective exposure~\cite{spohr2017fake}. Our model is agnostic to the attribution of blame for selective exposure and is based only on the assumption of confirmation bias.

The potentially negative influence of social media on the quality of democratic decisions has also been studied~\cite{tucker2018social}. Threats to democracy due to the mutually-boosting effects of disinformation and polarization have been recently voiced~\cite{sunstein2018republic}. It was observed that exposure to more information did not necessarily improve the quality of political discourse~\cite{dimitrova2014effects}. 

This work is the first that develops an analytical model to understand how the sheer {\em volume of information\/} affects polarization, given (i) democratized production, (ii) the existence of confirmation bias (and therefore selective exposure)~\cite{nickerson1998confirmation}, and (iii) consumer preference to outlying content~\cite{lamberson2018model,varshney2019must}. A simplified model of social influence is also taken into account. 

We are also novel in adoption a diffusion-drift model, where the conflicting effects of belief drift (due to selective exposure) and diffusion (due to other random effects and chance encounters) are considered in modeling the impact of information (and disinformation) volume on the motion of nodes in the belief space, as well as on resulting polarization. The model is based on the Fokker-Planck equation~\cite{risken1996fokker}, traditionally used for modeling diffusion. This equation has a history of use in social contexts. In fact, a variation was proposed as a foundation for studying human behavioral models, subjected to forces in social fields~\cite{helbing1993boltzmann}. The Fokker-Planck equation was also used for modeling social differentiation and division of labor in learning systems~\cite{chen1987origin}, interactions between different social groups~\cite{weidlich1973fokker}, and the emergence of polarization when individuals in social groups make choices between discrete opposing alternatives~\cite{weidlich1971statistical}. In this paper, we use it to show the existence of a paradox: the sheer availability of more information increases polarization.

This work (and the key observation above) is related to previous results on the effect of network connectivity and homophilly on the probability of holding group-related beliefs~\cite{bienenstock1990effect}. 
The correlation between increased access and polarization was also observed in other contexts. For example, as mentioned earlier in the paper, it was shown that the creation of the interstate highway system in the US increased geographic polarization~\cite{nall2015political}, as the ease of commute facilitated urban sprawl and the rise of internally-homogeneous geographic neighborhoods (in an analogy with social echo-chambers). 
In urban contexts, where the impact of total size was studied the most, it was also shown that urban growth is correlated with increased (price) polarization~\cite{antoniucci2018social}. Our model suggests that similar observations regarding the impact of connectivity (or access) and volume seem to apply to the information landscape as well.

\section{Conclusions}
The paper presented a social phenomenon caused by the age of democratized access; namely, growing ideological fragmentation exacerbated by information overload. A diffusion-drift model of this phenomenon was proposed. The model suggests that overload, in the presence of confirmation bias and preference for more outlying content, contributes to growing polarization. The paper is call for solutions that may ameliorate this effect. Technical advances in misinformation detection, securing content provenance, and unbiased information summarization can help mitigate the modeled effects. Issues related to incentives, ethics, and technology persuasion remain to be resolved.

\bibliographystyle{IEEEtran}
\bibliography{main,polarization}

\section*{Appendix A}
The Fokker-Planck equation of motion~\cite{fokker_planck} is the following:
\begin{equation}
    dX_t = \tilde{\mu}(X_t,t)dt + \sigma(X_t,t)dW\,,
\end{equation}
which can be written as
\begin{equation}
    \frac{dX_t}{dt} = \tilde{\mu}(X_t,t) + \sigma(X,t) \frac{dW}{dt}.
\end{equation}
Here $\tilde{\mu}(X_t,t)$ can be any function, and $X_t$ is the random variable.
From~\cite{fokker_planck}, we can see that the Fokker-Planck equation can be obtained as the following equation:
(Note: $p(x,t)$ here is probability density which is analogy to our population density $\rho(x,t)$.)
\begin{equation}
\label{eq:fp}
    \frac{\partial}{\partial t}p(x,t) = -\frac{\partial}{\partial x}
    \left[\tilde{\mu}(x,t)p(x,t) \right] + \frac{\partial^2}{\partial x^2}
    \left[\tilde{D}(x,t)p(x,t) \right]\,,
\end{equation}
which is the second equation on~\cite{fokker_planck}, and:
\begin{equation}
    \tilde{D}(x,t) = \frac{\sigma^2(x,t)}{2}.
\end{equation}
From the continuity equation:
\begin{equation}
\label{eq:con}
     \frac{\partial}{\partial t}p(x,t) = - \frac{\partial}{\partial x}j(x,t)\,,
\end{equation}
combining Eq.~\eqref{eq:con} and Eq.~\eqref{eq:fp}, one can see that the current reads:
\begin{equation}
    \label{eq:j2}
    j = 
    \tilde{\mu}(x,t)p(x,t) - \frac{\partial}{\partial x}
    \left[\tilde{D}(x,t)p(x,t) \right]\,.
\end{equation}

Now, let us compare with the equation of motion of our model:
\begin{equation}
        \frac{d x_{i,t}}{dt} \simeq \mu (\ln \eta_t(x_{i,t}))^\prime  + \mu \rho^{\prime}(x_{i,t},t)/\rho(x_{i,t},t)
    +\sigma\frac{dW}{dt}\,,
\end{equation}
which can be written as: 
\begin{equation}
    \frac{d x_{i,t}}{dt} \simeq \tilde{\mu}(x_{i,t},t)\,,
    +\sigma\frac{dW}{dt}
\end{equation}
with
\begin{equation}
    \tilde{\mu}(x_{i,t},t) = \mu (\ln \eta_t(x_{i,t}))^\prime  + \mu \rho^{\prime}(x_{i,t},t)/\rho(x_{i,t},t)\,.
\end{equation}
Therefore our current formula can be derived from Eq.~\eqref{eq:j2} as
\begin{align}
    j(x,t) &=  \tilde{\mu}(x,t) \rho(x,t) - \frac{\partial}{\partial x}
    \left[\frac{\sigma^2}{2}\rho(x,t) \right]\\
    &= \mu \rho(x,t)\frac{\partial}{\partial x}(\ln \eta_t(x))  + \mu \frac{\partial}{\partial x}\rho(x,t) 
    -\frac{\sigma^2}{2} \frac{\partial}{\partial x}\rho(x,t)\\
    &= -(\frac{\sigma^2}{2}-\mu)\frac{\partial}{\partial x}\rho(x,t) 
    + \mu\rho(x,t) \frac{\partial}{\partial x}\left(\ln \eta_0(x) \right)
    + \mu\rho(x,t) \frac{\partial}{\partial x}\left(\lambda \rho(x,t) \right)\\
    &=-D\frac{\partial}{\partial x}\rho(x,t) 
    - \mu\rho(x,t) \frac{\partial}{\partial x}V(x)
    -\mu\rho(x,t)\frac{\partial}{\partial x}\left(g\rho(x,t)\right)\,,
\end{align}
with $D = \frac{\sigma^2}{2}-\mu$, $V(x) = -\ln \eta_0(x)$, and $g=-\lambda$.
$\blacksquare$

\end{document}